\documentclass[preprint,showpacs, aps]{revtex4}
\usepackage{graphicx}% Include figure files

\begin{document}
%\input  epsf
%\draft

\title{Localized magnetic states in biased bilayer and trilayer graphene}
\author {Kai-He Ding$^1$, Zhen-Gang Zhu$^2$, and Jamal Berakdar$^2$}
\affiliation{$^1$Department of Physics and Electronic Science,
 Changsha University of Science and Technology,
 Changsha,410076, China\\
 $^2$Institut f\"{u}r Physik Martin-Luther-Universit\"{a}t
Halle-Wittenberg, Nanotechnikum-Weinberg, Heinrich-Damerow-Strasse 4
D - 06120 Halle (Saale), Germany}

%\date{\today }
\begin{abstract}
We study the localized magnetic states of impurity in biased bilayer
and trilayer graphene. It is found that the magnetic boundary for
bilayer and trilayer graphene presents the mixing features of Dirac
and conventional fermion. For zero gate bias, as the impurity energy
approaches the Dirac point, the impurity magnetization region
diminishes for bilayer and trilayer graphene. When a gate bias is
applied, the dependence of impurity magnetic states on the impurity
energy exhibits a different behavior for bilayer and trilayer
graphene due to the opening of a gap between the valence  and the
conduction  band in the bilayer graphene with the  gate bias applied.
The magnetic moment and the corresponding magnetic transition of the
impurity in bilayer graphene are also investigated.

\end{abstract}

 \maketitle

 \section{Introduction}
The intense  research currently devoted to graphene, a  two-dimensional carbon honeycomb lattice,
has uncovered a wealth of
fascinating  properties  such as the anomalous quantized
Hall effect,  the absence of the weak localization and existence of the
minimal
conductivity\cite{novoselov2,zhang,geim,castrocond,novoselovsci2004}.
Graphene has a high mobility, its  carrier density is controllable by an applied
gate voltage\cite{zhang} and a spin-orbit
interaction\cite{berger,huertas,kane1,yao,ding1}.

 Graphene structures  have been  the focus of much interest
\cite{Ponomarenko,Pedersenprl2008,Novoselovnat2006,Castroprl2008,Castroprl20082,Bostwicknjp,han2007,Wimmerprl2008,Wangarxiv,guonano2007}.
In particular, adatoms may be positioned on graphene by current
nanotechnology\cite{eiglernature1990}, rendering the study and
manipulation of local electronic properties. \emph{Ab initio}
calculations for transition metal adatoms\cite{duffyprb1998} show
a tendency to the formation of local magnetic moments. Recently
Uchoa \emph{et al.}\cite{Uchoaprl2008} examined the condition  for
the emergence of localized magnetic moment on adatoms with inner
shell electrons on a single layer graphene. It is found that the
impurity magnetization boundary exhibits anomalous
characteristics. In contrast to the case of an impurity in an
ordinary metal, the impurity can magnetize for any small charging
energy due to the low density of state(DOS) at the Dirac point. On
the other hand, detailed experimental studies \cite{Bostwicknjp}
on multi-layer graphene showed a marked modification of the
electronic structure with the number of layers. Hence, we expect
\cite{ding2} a qualitative difference in the magnetic properties
of the adatoms on multilayer graphene; an issue which we address
here by inspecting
  the localized magnetic state of  an impurity in
a biased bilayer and trilayer graphene.
We find that the size of the magnetic region decreases rapidly
compared with that in monolayer graphene, the impurity can
magnetize even when the energy of the doubly occupied state is
below the Fermi level, and the impurity magnetization region is
asymmetric due to the special
  nature of the quasiparticles having mixed features of Dirac
and conventional fermions. When a gate bias is applied, the
dependence of the impurity magnetic states on the impurity energy
for a bilayer graphene exhibits a different behavior from that for
a trilayer graphene due to the opening of a  gate-induced gap
between  the valence and the conduction band in the bilayer
graphene. Calculating the occupation of the impurity level and the
susceptibility in the bilayer graphene we show that the magnetic
moment decreases with increasing the inter-layer coupling.
% For an impurity energy  close to Dirac point,
%the local magnetic moment decreases for the unbiased bilayer, while
%it exhibits a nonmonotonic dependence on the impurity energy for the
%biased bilayer, which is also related to the band gap induced by the
%gate bias.
%
%The paper is organized as follows. In Sec.II, we derive the model
%for the biased bilayer graphene with an impurity, and the formula of
%the occupation of the impurity level in this model. In Sec.III, we
%derive the formula of the occupation of the impurity in the biased
%trilayer graphene. In Sec.IV, the corresponding numerical results
%are given. Finally, a summary is presented.
%
\section{Bilayer graphene}
\begin{figure}[h]
\includegraphics[width=0.8\columnwidth ]{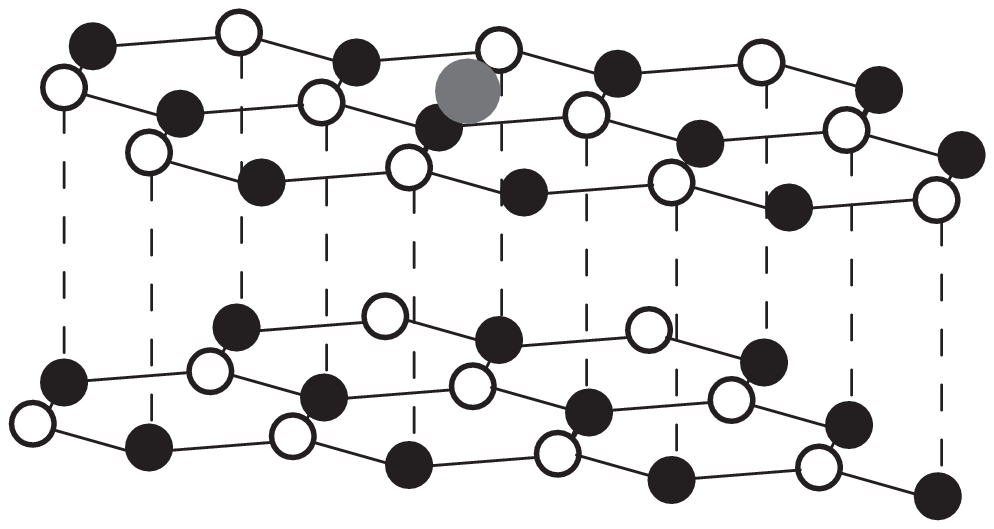}
\caption{Schematic diagram of the lattice structure of the bilayer
graphene with an impurity atom.}\label{fig1}
%\end{center}
\end{figure}

Fig.\ref{fig1} shows the lattice structure of the bilayer graphene with
the  adatom. The
inter-layer stacking is assumed to be the Bernal order where the top layer
has its A sublattice atop the sublattice B of the bottom layer.
The bias voltage $V$ is applied across the  layers. The system
Hamiltonian
\begin{equation}
H=H_{TB}+H_i+H_f
\end{equation}
contains the graphene bilayer  term $H_{TB}$, which in a
tight-binding approximation reads
\begin{equation}
H_{TB}=\sum\limits_{l=1}^2 H_l+H_T+H_V,
\end{equation}
with
\begin{equation}
H_l=-t\sum\limits_{\langle
i,j\rangle\sigma}[a_{l\sigma}^\dag(\mathbf{R}_i)b_{l\sigma}(\mathbf{R}_j)+b_{l\sigma}^\dag(\mathbf{R}_j)
a_{l\sigma}(\mathbf{R}_i)],\label{h1}
\end{equation}
\begin{equation}
H_T=-t_p\sum\limits_{i,\sigma}[a_{1\sigma}^\dag(\mathbf{R}_i)b_{2\sigma}(\mathbf{R}_i)
+b_{2\sigma}^\dag(\mathbf{R}_i)
a_{1\sigma}(\mathbf{R}_i)],\label{h2}
\end{equation}
\begin{equation}
H_V=\frac{V}{2}\sum_{i\sigma}[a_{1\sigma}^\dag(\mathbf{R}_i)a_{1\sigma}(\mathbf{R}_i)+b_{1\sigma}^\dag(\mathbf{R}_i)
b_{1\sigma}(\mathbf{R}_i)
-a_{2\sigma}^\dag(\mathbf{R}_i)a_{2\sigma}(\mathbf{R}_i)-b_{2\sigma}^\dag(\mathbf{R}_i).
b_{2\sigma}(\mathbf{R}_i)],\label{h3}
\end{equation}
The operator  $a_{l\sigma}(\mathbf{R}_i)\; (b_{l\sigma}(\mathbf{R}_i))$
annihilates a state with a spin $\sigma$ at the position
$\mathbf{R}_i$ on the sublattice A(B) of the $l$ plane. $t$ is the
nearest neighbour in-plane hopping energy, $t_p$ is the
inter-layer hopping energy. For the hybridization with the localized
impurity states  we write
% at a given site, such as sublattice B of the
% plane 1, is given by
\begin{equation}
H_i=V_f\sum\limits_{\sigma}[f_\sigma^\dag
b_{1\sigma}(0)+b_{1\sigma}^\dag(0) f_\sigma],
\end{equation}
where $f_\sigma (f_\sigma^\dag)$ is the  annihilation
(creation) operator of a state with a spin $\sigma$ at the
impurity, and $V_f$ is the hybridization strength. In the momentum
space we have
\begin{equation}
H_l=-t\sum\limits_{\mathbf{k}\sigma} [\phi(\mathbf{k})
a_{l\mathbf{k}\sigma}^\dag
b_{l\mathbf{k}\sigma}+\phi^*(\mathbf{k})b_{l\mathbf{k}\sigma}^\dag
a_{l\mathbf{k}\sigma}],\label{mh1}
\end{equation}
\begin{equation}
H_T=t_p\sum\limits_{\mathbf{k},\sigma}[a_{1\mathbf{k}\sigma}^\dag
b_{2\mathbf{k}\sigma} +b_{2\mathbf{k}\sigma}^\dag
a_{1\mathbf{k}\sigma}],\label{mh3}
\end{equation}
\begin{equation}
H_V=\frac{V}{2}\sum_{\mathbf{k}\sigma} [a_{1\mathbf{k}\sigma}^\dag
a_{1\mathbf{k}\sigma}+b_{1\mathbf{k}\sigma}^\dag
b_{1\mathbf{k}\sigma} -a_{2\mathbf{k}\sigma}^\dag
a_{2\mathbf{k}\sigma}-b_{2\mathbf{k}\sigma}^\dag
b_{2\mathbf{k}\sigma}],\label{mh2}
\end{equation}
\begin{equation}
H_i=\frac{V_f}{\sqrt{N}}\sum\limits_{\mathbf{k}\sigma}(f_\sigma^\dag
b_{1\mathbf{k}\sigma}+b_{1\mathbf{k}\sigma}^\dag f_\sigma),
\end{equation}
where $\phi(\mathbf{q})=-t\sum\limits_{i=1}^3
e^{i\mathbf{q}\cdot\mathbf{\delta_i}}$ with
$\delta_1=\frac{a}{2}(1,\sqrt{3},0),
\delta_2=\frac{a}{2}(1,-\sqrt{3},0),\delta_3=a(1,0,0)$ (here $a$
is the lattice spacing), and $N$ is the number of sites on
sublattice B of plane 1.
 Diagonalizing   $H_{TB}$ we find the spectrum
\begin{equation}
E_{\pm\pm}(\mathbf{k})=\pm\sqrt{\epsilon_k^2
+\frac{t_p^2}{2}+\frac{V^2}{4}\pm\sqrt{\frac{t_p^4}{4}+(t_p^2+V^2)\epsilon_k^2}},
\end{equation}
where $\epsilon_k=\pm |\phi(\mathbf{k})|$ is linearizable
 around the $\mathbf{K}$
points of the Brillouin zone by $\epsilon_k=\pm v_F|\mathbf{k}|$
where  $v_F=3ta/2$ is the Fermi velocity.
The impurity is described by Hamiltonian $H_f$ with
\begin{equation}
H_f=\sum\limits_\sigma \varepsilon_0 f^\dag_\sigma f_\sigma+U
n_\uparrow n_\downarrow,\label{hf1}
\end{equation}
where $n_\sigma=f_\sigma^\dag f_\sigma$ is the occupation number
operator, $\varepsilon_0$ is the single electron energy at the
impurity. The Coulomb interaction is included as a finite
Anderson term $U$. For simplicity, we adopt a mean field approximation
to the electronic correlations at the impurity, $Un_\uparrow
n_\downarrow \simeq U\sum\limits_\sigma \langle
n_{\overline{\sigma}}\rangle f_\sigma^\dag f_\sigma-U\langle
n_\uparrow\rangle \langle n_\downarrow\rangle$. The impurity
Hamiltonian is rewritten as $H_f=\sum\limits_\sigma
\varepsilon_\sigma f^\dag_\sigma f_\sigma$ with
$\varepsilon_\sigma=\varepsilon_0+U\langle
n_{\overline{\sigma}}\rangle$.
To investigate the localized magnetic states, we
calculate the occupation number of the electrons of a given spin
$\sigma$ at the impurity. At low temperatures all the
states below the Fermi level $\mu$ are completely occupied and the
occupation of the impurity is determined by
\begin{equation}
\langle n_\sigma\rangle=\int_{-\infty}^\mu d\omega
\rho_f^\sigma(\omega),\label{nsigma}
\end{equation}
where $\rho_f^\sigma(\omega)$ is DOS at the impurity level. We infer it
from the retarded Green's function
\begin{equation}
G_{f}^{r,\sigma\sigma'}(t-t')=-i\theta(t-t')\langle\{f_\sigma(t),f_{\sigma'}^\dag(t')\}\rangle.
\label{gff1}
\end{equation}
By the standard equation of motion, we can derive
\begin{equation}
 G_{f}^{r,\sigma\sigma'}(\omega)=\frac{\delta_{\sigma\sigma'}}{\omega-\varepsilon_\sigma-\Sigma_{f}^r(\omega)+i\eta},
 \label{gfr1}
\end{equation}
where
\begin{equation}
\begin{array}{cll}
\Sigma_{f}^r(\omega)
&=&\frac{V_f^2}{N}\sum\limits_\mathbf{k}\frac{-(\omega-\frac{V}{2})v_F^2|\mathbf{k}|^2+(\omega-\frac{V}{2})
(\omega+\frac{V}{2})^2- t_p^2(\omega+\frac{V}{2})}
{v_F^4|\mathbf{k}|^4-2(\omega^2+\frac{V^2}{4})v_F^2|\mathbf{k}|^2+
(\omega^2-\frac{V^2}{2})^2- t_p^2(\omega^2-\frac{V^2}{4})}.
\end{array}\label{sigmaf1}
\end{equation}
Introducing a high-energy cutoff $D$ of the graphene bandwidth,
we obtain  for
$\omega^2\geq\frac{1}{4}\frac{t_p^2V^2}{t_p^2+V^2}$,
\begin{equation}
\begin{array}{cll}
\Sigma_{f}^r(\omega) &=&\frac{V_f^2}{D^2}\{
\frac{V\omega^2-(V^2/2+t_p^2)\omega-Vt_p^2/2}{\sqrt{4(V^2+t_p^2)\omega^2-t_p^2V^2}}
\ln|\frac{(D^2-x_1)x_2}{(D^2-x_2)x_1}|-\frac{\omega-V/2}{2}\ln
|\frac{(D^2-x_1)(D^2-x_2)}{x_1x_2}|\}\\
&&+i \frac{\pi V_f^2}{D^2}
\{\frac{V\omega^2-(V^2/2+t_p^2)\omega-Vt_p^2/2}
{\sqrt{4(V^2+t_p^2)\omega^2-t_p^2V^2}} [\text{sgn}(\frac{d
x_1}{d\omega})\theta(0<x_1<D^2)
-\text{sgn}(\frac{d x_2}{d\omega})\theta(0<x_2<D^2)]\\
&&-\frac{\omega-V/2}{2}[\text{sgn}(\frac{d
x_1}{d\omega})\theta(0<x_1<D^2)+\text{sgn}(\frac{d
x_2}{d\omega})\theta(0<x_2<D^2)]\} ,\label{sigmaf}
\end{array}
\end{equation}
where $\theta(x)$ is the step function, and
\begin{equation}
x_{1,2}=(\omega^2+\frac{V^2}{4})\pm\frac{1}{2}\sqrt{4(V^2+t_p^2)\omega^2-t_p^2V^2}.
\end{equation}
For $\omega^2<\frac{1}{4}\frac{t_p^2V^2}{t_p^2+V^2}$,
\begin{equation}
\begin{array}{cll}
\Sigma_{f}^r(\omega)
&=&\frac{V_f^2}{D^2}[-\frac{\omega-V/2}{2}\ln|\frac{D^4-2(\omega^2+V^2/4)D^2+
(\omega^2-V^2/2)^2- t_p^2(\omega^2-V^2/4)}{(\omega^2-V^2/2)^2-
t_p^2(\omega^2-V^2/4)}|\\
&&+\frac{V\omega^2-(V^2/2+t_p^2)\omega-Vt_p^2/2}{\sqrt{t_p^2V^2/4-(V^2+t_p^2)\omega^2}}
(\arctan\frac{D^2-\omega^2-V^2/4}{\sqrt{t_p^2V^2/4-(V^2+t_p^2)\omega^2}}
+\arctan\frac{\omega^2+V^2/4}{\sqrt{t_p^2V^2/4-(V^2+t_p^2)\omega^2}})].\\
\end{array}
\end{equation}
The summation over
$\mathbf{k}$   in Eq.(\ref{sigmaf1}) is accurate for $\omega\ll D$ by
ensuring  the conservation of the total number of states in the
Brillouin zone according to the Debye's prescription. Substituting
$\Sigma_{f}^r(\omega)$ into Eq.(\ref{gfr1}), the retarded Green's
function $G_f^{r,\sigma\sigma}(\omega)$ can be obtained. Note,
 the determination of $\langle
n_\sigma\rangle$ in Eq.(\ref{nsigma}) entails  a  self-consistent
calculation of DOS at the impurity level via the relation
$\rho_f^\sigma(\omega)=-\frac{1}{\pi}\text{Im}G_{f}^{r,\sigma\sigma}(\omega)
$.  When $t_p=V=0$, our present results  reduce to those of
Ref.\cite{Uchoaprl2008}.

\section{Trilayer graphene}
The Hamiltonian for trilayer graphene contains
 a coupling the B atom of the second layer to
the A atom of the third  layer according to the conventional Bernal-type
stacking order. Similar to the bilayer graphene case  we find
for the impurity Green's function
\begin{equation}
G_{f}^{r,\sigma\sigma'}(\omega)=\frac{\delta_{\sigma\sigma'}}{\omega-\epsilon_\sigma-\Sigma^r+i\eta},\label{2Lfeq02}
\end{equation}
where
\begin{equation}
\begin{array}{cll}
\Sigma^r
&=&-\frac{V_f^2}{N}\sum\limits_{\mathbf{k}}\frac{A_1v_F^4|\mathbf{k}|^4+B_1v_F^2|\mathbf{k}|^2+C_1}
{v_F^6|\mathbf{k}|^6+B_2v_F^4|\mathbf{k}|^4+C_2v_F^2|\mathbf{k}|^2+D_2}
\end{array}\label{sigmar}
\end{equation}
with $A_1=\omega-\frac{V}{2}$, $B_1=t_p^2\omega
-(\omega-\frac{V}{2})[\omega^2+(\omega+\frac{V}{2})^2]$,
$C_1=\omega^2(\omega-\frac{V}{2})(\omega+\frac{V}{2})^2
-2t_p^2\omega^2(\omega+\frac{V}{2})$,
$B_2=-3\omega^2-\frac{V^2}{2}$, $C_2=-2t_p^2\omega^2
+3\omega^4+\frac{V^4}{16}$, $D_2=-\omega^2(\omega^2-\frac{V^2}{4})^2
+2t_p^2\omega^2(\omega^2-\frac{V^2}{4})$.
Performing  the summation over
$\mathbf{k}$ in Eq.(\ref{sigmar}) as Eq.(\ref{sigmaf1})
%can be also
%performed by taking the continuum limit with the introducing of the
%cutoff $D$ and expanding the integrand in terms of the partial
%fractions. The final results are given
we find
for  $\Delta=(2B_2^3
-9B_2C_2+27D_2)^2+4(-B_2^2+3C_2)^3\geq 0$ the result
\begin{equation}
\begin{array}{cll}
\Sigma^r &=&-\frac{V_f^2}{D^2} \{[A_1+\frac{A_1(x_2+x_3)x_1+B_1x_1
-A_1x_2x_3+C_1}{(x_2-x_1)(x_3-x_1)}]\ln|\frac{D^2-x_1}{x_1}|\\
&&+\frac{A_1(x_2+x_3)+B_1}{\sqrt{x_2x_3-(x_2+x_3)^2/4}}
(\arctan\frac{D^2-(x_2+x_3)/2}{\sqrt{x_2x_3-(x_2+x_3)^2/4}}+\arctan\frac{(x_2+x_3)/2}{\sqrt{x_2x_3-(x_2+x_3)^2/4}})\\
&&+\frac{A_1(x_2+x_3)x_1+B_1x_1
-A_1x_2x_3+C_1}{(x_2-x_1)(x_3-x_1)}[\frac{-1}{2}\ln\frac{D^2-(x_2+x_3)+x_2x_3}{x_2x_3}\\
&&+\frac{(x_2+x_3)/2-x_1}{2\sqrt{x_2x_3-(x_2+x_3)^2/4}}
(\arctan\frac{D^2-(x_2+x_3)/2}{\sqrt{x_2x_3-(x_2+x_3)^2/4}}+\arctan\frac{(x_2+x_3)/2}{\sqrt{x_2x_3-(x_2+x_3)^2/4}})]\}\\

&&-i\text{sgn}(\frac{dx_1}{d\omega})\theta(0<x_1<D^2)\frac{\pi
V_f^2}{D^2} [A_1+\frac{A_1(x_2+x_3)x_1+B_1x_1
-A_1x_2x_3+C_1}{(x_2-x_1)(x_3-x_1)}],
\end{array}\label{tsigma1}
\end{equation}
where
\begin{equation}
\begin{array}{cll}
x_1 &=&-\frac{B_2}{3}+\frac{1}{2^{1/3}}\frac{1}{3}\{-2B_2^3
+9B_2C_2-27D_2+\sqrt{(2B_2^3
-9B_2C_2+27D_2)^2+4(-B_2^2+3C_2)^3}\}^{\frac{1}{3}}\\
&&+\frac{1}{2^{1/3}3}\{-2B_2^3 +9B_2C_2-27D_2-\sqrt{(2B_2^3
-9B_2C_2+27D_2)^2+4(-B_2^2+3C_2)^3}\}^{\frac{1}{3}},
\end{array}
\end{equation}
\begin{equation}
\begin{array}{cll}
x_{2,3}&=&-\frac{B_2}{3}+\frac{-\frac{1}{2}-i\frac{\sqrt{3}}{2}}{3}\frac{1}{2^{1/3}}\{-2B_2^3
+9B_2C_2-27D_2\pm\sqrt{(2B_2^3
-9B_2C_2+27D_2)^2+4(-B_2^2+3C_2)^3}\}^{\frac{1}{3}}\\
&&+\frac{-\frac{1}{2}+i\frac{\sqrt{3}}{2}}{3}\frac{1}{2^{1/3}}\{-2B_2^3
+9B_2C_2-27D_2\mp\sqrt{(2B_2^3
-9B_2C_2+27D_2)^2+4(-B_2^2+3C_2)^3}\}^{\frac{1}{3}}.
\end{array}
\end{equation}
For $\Delta<0$,
\begin{equation}
\begin{array}{cll}
&&\Sigma^r\\
&=&-\frac{V_f^2}{D^2}\{
[A_1+\frac{[A_1(x_2+x_3)+B_1]x_1-A_1x_2x_3+C_1}
{(x_2-x_1)(x_3-x_1)}]\ln|\frac{D^2-x_1}{x_1}|\\
&&-
[\frac{A_1(x_2+x_3)+B_1}{x_3-x_2}+\frac{[A_1(x_2+x_3)+B_1]x_1-A_1x_2x_3+C_1}
{(x_2-x_1)(x_3-x_2)}]\ln|\frac{D^2-x_2}{x_2}|\\
&&+
[\frac{A_1(x_2+x_3)+B_1}{x_3-x_2}+\frac{[A_1(x_2+x_3)+B_1]x_1-A_1x_2x_3+C_1}
{(x_2-x_1)(x_3-x_2)}-\frac{[A_1(x_2+x_3)+B_1]x_1-A_1x_2x_3+C_1}
{(x_2-x_1)(x_3-x_1)}]\ln|\frac{D^2-x_3}{x_3}|\}\\
&&-i\frac{\pi
V_f^2}{D^2}\{\text{sgn}(\frac{dx_1}{d\omega})\theta(0<x_1<D^2)
[A_1+\frac{[A_1(x_2+x_3)+B_1]x_1-A_1x_2x_3+C_1}
{(x_2-x_1)(x_3-x_1)}]\\
&&-\text{sgn}(\frac{dx_2}{d\omega})\theta(0<x_2<D^2)
[\frac{A_1(x_2+x_3)+B_1}{x_3-x_2}+\frac{[A_1(x_2+x_3)+B_1]x_1-A_1x_2x_3+C_1}
{(x_2-x_1)(x_3-x_2)}]\\
&&+\text{sgn}(\frac{dx_3}{d\omega})\theta(0<x_3<D^2)
[\frac{A_1(x_2+x_3)+B_1}{x_3-x_2}+\frac{[A_1(x_2+x_3)+B_1]x_1-A_1x_2x_3+C_1}
{(x_2-x_1)(x_3-x_2)}-\frac{[A_1(x_2+x_3)+B_1]x_1-A_1x_2x_3+C_1}
{(x_2-x_1)(x_3-x_1)}]\},
\end{array}\label{tsigma2}
\end{equation}
where
\begin{equation}
\begin{array}{cll}
x_1 &=&-\frac{B_2}{3}+\frac{2\sqrt{B_2^2-3C_2}}{3}\cos(
\frac{\arccos T}{3}),\ \ \ x_2
=-\frac{B_2}{3}+\frac{2\sqrt{B_2^2-3C_2}}{3}\cos( \frac{2\pi+\arccos
T}{3}),
\end{array}
\end{equation}
\begin{equation}
\begin{array}{cll}
 x_3 =-\frac{B_2}{3}+\frac{2\sqrt{B_2^2-3C_2}}{3}\cos(
\frac{4\pi+\arccos T}{3}),\ \ \ T=-\frac{2(B_2^2-3C_2)B_2
-3(B_2C_2-9D_2)}{2(B_2^2-3C_2)^{\frac{3}{2}}}.
\end{array}
\end{equation}
Substituting Eqs.(\ref{tsigma1}) and (\ref{tsigma2}) in
Eq.(\ref{2Lfeq02}), we can derive self-consistently the occupation on the
 impurity for case of a trilayer graphene.
\section{Numerical analysis}
\begin{figure}[h]
\includegraphics[width=0.8\columnwidth ]{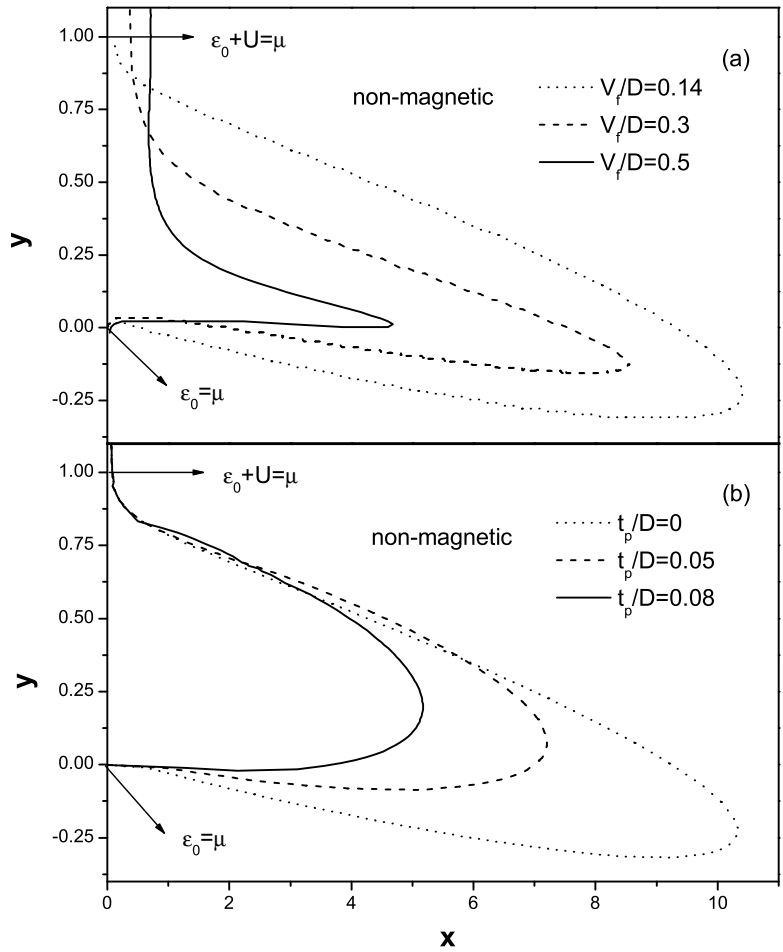}
\caption{Regions of the magnetic and the nonmagnetic phase for the
bilayer graphene. The boundary line gives $y$ as a function of $x$ (defined in eq.(\ref{eq:xy}))
at $t_p/D=0$ for the different $V_f/D$ (a), and at $V_f/D=0.14$ for
 different $t_p/D$ (b). The other parameters are
$\varepsilon_0/D=0.029$ and $V/D=0$.}\label{fig2}
\end{figure}

\begin{figure}[h]
\includegraphics[width=0.8\columnwidth ]{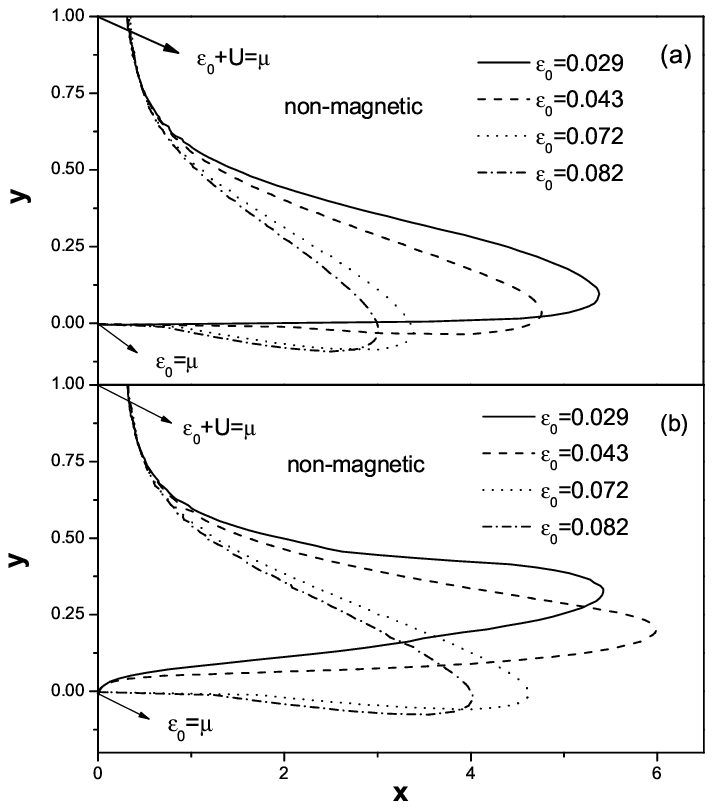}
\caption{Regions of magnetic and non-magnetic phase for the
bilayer graphene. The boundary line gives $y$ as a function of $x$
for the different $\varepsilon_0/D$ at $V/D=0$ (a) and at $V/D=0.05$
(b), where $V_f/D=0.3$ and $t_p/D=0.05$. }\label{fig3}
\end{figure}

 From the occupation of the two spin
channel on the impurity we conclude on the formation
of localized magnetic   moment  whenever
$n_\uparrow\neq n_\downarrow$. For a detailed study conventionally,
one   introduces the dimensionless
parameters
\begin{equation}
 x=D\Gamma/U\;  \mbox{ and }\; y=(\mu-\varepsilon_0)/U
 \;  \mbox{ with }\; \Gamma=\pi V_f^2/D^2.\label{eq:xy}
\end{equation}
 The transition curves from the magnetic to
the non-magnetic behavior as a function of the parameters $x$ and
$y$ for the different hybridization and inter-layer coupling in the
bilayer graphene are shown in Fig.\ref{fig2}. For $t_p=V=0$,
our results reduce to those of  Ref.\cite{Uchoaprl2008}: The magnetic
boundary exhibits an asymmetry around $y=0.5$, and can even cross when
line $y=1$. The magnetic region shrinks in the $x$ direction with
 the hybridization  $V_f$ is increased; for $y$ close to $1$ (cf. eq.(\ref{eq:xy})),
  the boundary line for magnetic transition  shifts away   from the
$y$ axis due to the increased influence of  graphene on the
impurity magnetization with  enhanced  hybridization. When
the inter-layer coupling $t_p$ is taken into account (see
Fig.\ref{fig2}(b)), the size of the magnetic region diminishes
rapidly, and for a large enough  $t_p$, the magnetic boundary
shrinks above the line $y=0$. However, the magnetic boundary does
not turn symmetric around $y=0.5$, and the above magnetic boundary
line crosses the line $y=1$. The origin of this phenomena lies  in the peculiar
nature of the quasiparticles in the bilayer graphene; they exhibits
 features akin both to Dirac and to conventional fermions.
The contribution of conventional fermions originates from   the
interlayer coupling that supports a metallic bilayer graphene
and results in effects as for a conventional  metallic host  on the magnetic
properties of the impurity. For large interlayer
coupling  we observe  therefore
 magnetic boundaries similar
an impurity  in an ordinary metal.
 (Fig.\ref{fig3}) shows  for a bilayer graphene
  the
boundary between magnetic and non-magnetic impurity states as a
function of the parameters $x$ and $y$ (eq.\ref{eq:xy}) for  different impurity
energy levels $\varepsilon_0$.
For $V=0$ the size of the magnetic region grows as
$\varepsilon_0$ approaches the energy of the Dirac point. This
behavior is reminiscent of the single layer of
graphene\cite{Uchoaprl2008}, and originates from the suppression of
the DOS around the impurity energy level. In contrast, for a
nonzero gate bias, when $\varepsilon_0$ is close to the Dirac point
from the positive energy side, the size of the region first increases
to the maximum, then decreases with decreasing $\varepsilon_0$, as
shown in Fig.\ref{fig3}(b). The explanation for this phenomenon is
as follows: the gate bias voltage gives rise to a finite electronic
gap between the conduction and the valence band, and induces a large
local DOS close to the gap edges\cite{Eduardo}. In particular, the
DOS may extend into the gap due to the influence of the
impurity\cite{Nilssonprl}. In this situation, the coupling between the
bath and the impurity is enhanced inside the gap as compared with the zero
bias case,  leading thus to the non-monotonic dependence of the size of
the region with $\varepsilon_0$.

\begin{figure}[h]
\includegraphics[width=0.8\columnwidth ]{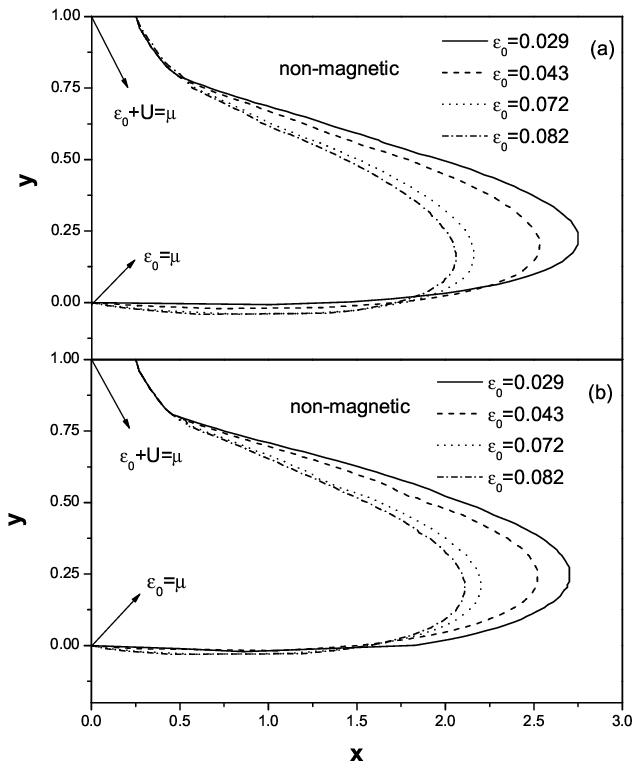}
\caption{Regions of magnetic and non-magnetic phase for the
trilayer graphene. The boundary line gives $y$ as a function of $x$
for the different $\varepsilon_0/D$ at $V/D=0$ (a) and at $V/D=0.05$
(b), where $V_f/D=0.2$ and $t_p/D=0.05$. }\label{fig4}
\end{figure}

Fig.\ref{fig4} shows the magnetic transition curve as a function of
the parameters $x$ and $y$ (eq.\ref{eq:xy})
 for  different $\varepsilon_0$ in the
trilayer graphene. For $V=0$,  phenomena such as the asymmetry around
the line $y=0.5$ and the crossing of the line $y=1$ in the magnetic
boundary suggest the existence of Dirac fermions in the trilayer
graphene. As $\varepsilon_0$ approaches the energy of the Dirac
point, the magnetization region of the impurity grows due to the two
almost-linear touched bands reminiscent of the bands in monolayer
graphene\cite{Bostwicknjp}. It is interesting to note that for
nonzero gate bias, the impurity magnetization region increases
monotonously when $\varepsilon_0$ is close to the Dirac point, which is
clearly different from that in the bilayer graphene. This behavior
stems from the fact that the gate bias can not destroy the
particle-hole degeneracy in the trilayer graphene\cite{Bostwicknjp}.

\begin{figure}[h]
\includegraphics[width=0.8\columnwidth ]{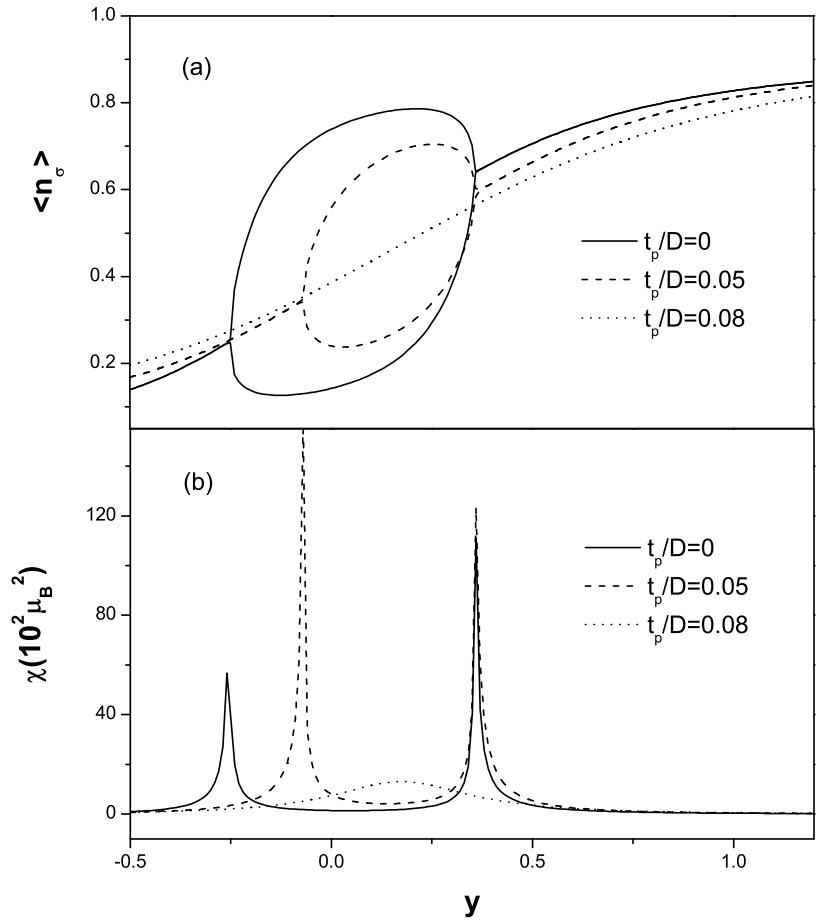}
\caption{The occupation of the impurity spin level and the magnetic
susceptibility in the bilayer graphene for the different $t_p/D$ at
 $x=6$.
 The other parameters are  those of
 Fig.\ref{fig2}(b).}\label{fig5}
\end{figure}

\begin{figure}[h]
\includegraphics[width=0.8\columnwidth ]{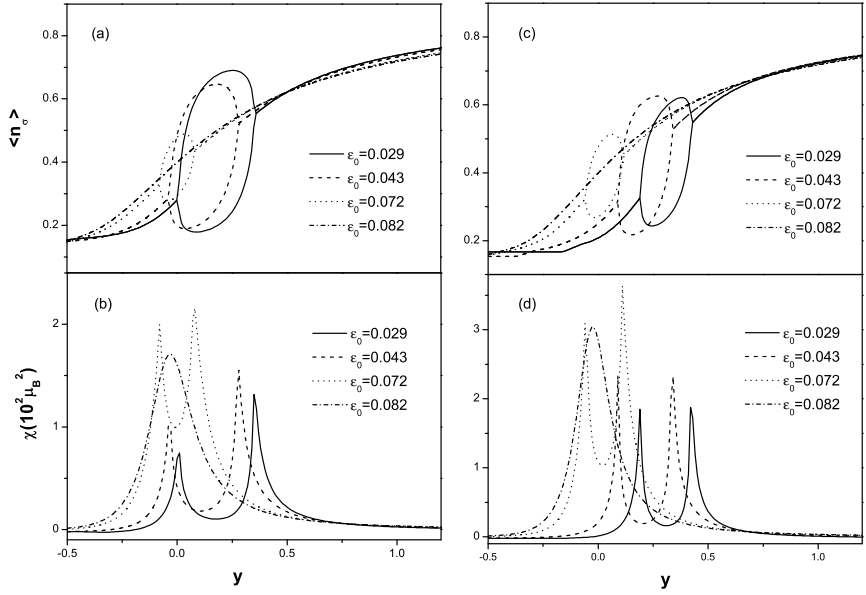}
\caption{The occupation of the impurity spin level and the magnetic
susceptibility in the bilayer graphene for  different
$\varepsilon_0/D$ at $V/D=0$ and $x=3.2$   (a)-(b), and at
$V/D=0.05$
 and $x=4.2$ (c)-(d).
 The other parameters are  the same as in
 Fig.\ref{fig3}}\label{fig6}
\end{figure}

To investigate the localized magnetic moment of the
impurity in the magnetic region and the magnetic transition we
 calculate the magnetic susceptibility.
 The energy of the impurity spin states
in a  magnetic field $B$ is $\varepsilon_\sigma=\varepsilon_0-\sigma\mu_B
B+Un_{\overline{\sigma}}$. The  magnetic
susceptibility of the impurity derives from
\begin{equation}
\chi=-\mu_B^2\sum\limits_\sigma \frac{d\langle
n_\sigma\rangle}{d\varepsilon_\sigma}\frac{1-U\frac{d\langle
n_{\overline{\sigma}}\rangle}{d\varepsilon_{\overline{\sigma}}}}
{1-U^2\frac{d\langle
n_{\overline{\sigma}}\rangle}{d\varepsilon_{\overline{\sigma}}}\frac{d\langle
n_\sigma\rangle}{d\varepsilon_\sigma} }.
\end{equation}
Fig.\ref{fig5} shows the occupation of the impurity spin level and the magnetic
susceptibility as a function of $y$ for the different inter-layer
coupling in a bilayer graphene. The occupation $\langle n_\sigma\rangle$ versus $y$
is a bubble that corresponds to the impurity magnetization.
The corresponding susceptibility exhibits two peaks at the
magnetization edge indicating the strength of the magnetic
transition. For $t_p=0$, a strong magnetic moment of $\sim 0.7\mu_B$
forms in almost the whole magnetic region. With increasing the
inter-layer coupling $t_p$, the magnetic bubble region diminishes
signalling the decrease of the magnetic moment of the impurity, and
the magnetic transition becomes very sharp. There is no localized
magnetic moment in the case of a sufficiently strong inter-layer
coupling. In this case, the magnetic boundary shrinks below the line
$x=6$ in the $x$ direction(see Fig.\ref{fig2}(b)).
 Fig.\ref{fig6} shows the occupation of the impurity level and
the magnetic susceptibility as a function of $y$ for the different
impurity energy level $\varepsilon_0$ in the bilayer graphene. The
corresponding magnetic boundaries are defined in Fig.\ref{fig3} (a)
and (b) respectively. For $V=0$, the magnetic bubble shifts towards
the $\langle n_\sigma\rangle$ axis, and decreases with increasing
$\varepsilon_0$. When $\varepsilon_0$ becomes  large enough, the
bubble vanishes, meaning  that the impurity loses  magnetism
 in this situation. For large $\varepsilon_0$
 the magnetic transition
 becomes very sharp. Inspecting Fig.\ref{fig6}(c) and (d) we find
 when the gate bias $V$ is applied, the magnetic bubble
shows a  non-monotonic dependence on $\varepsilon_0$, while the
magnetic transition becomes very sharp with increasing
$\varepsilon_0$. Since the magnetic boundary line shrinks in the
left hand side of the line $x=4.2$ at $\varepsilon_0/D=0.082$(see
Fig.\ref{fig3}(b)), the impurity remains non-magnetic for any $y$,
i.e. $n_\uparrow=n_\downarrow$, as shown in Fig.\ref{fig6} (c).

\section{conclusions}
Summarizing, we studied the localized magnetic states of an impurity in
 biased bilayer and trilayer graphene. We find that the size of the magnetic region decreases
rapidly compared with that in monolayer graphene, the impurity can
magnetize even when the energy of the doubly occupied state is
below the Fermi level, and the impurity magnetization region has a
different shape. We can trace this behaviour back to the special
nature of quasiparticles. When a gate bias is applied, the
dependence of the impurity magnetic states on the impurity energy
for the bilayer graphene shows a behavior different from that for
a trilayer graphene due to the opening of a gap between the
valence and the conduction  band in the bilayer graphene.
 Correspondingly, the magnetic moment of the
impurity versus the impurity energy in the bilayer graphene is
affected strongly by the band gap induced by the gate bias.

\begin{acknowledgments}{ The work of K.H.D. was supported by the Natural
Science Foundation of Hunan Province, China (Grant No. 08JJ4002 ),
the National Natural Science Foundation of China (Grant No.
60771059), and Education Department of Hunan Province, China. J.B.
and Z.H.Z. were supported by the cluster of excellence
"Nanostructured Materials" of the state Saxony-Anhalt. }
\end{acknowledgments}

\newpage

\end{document}